\documentclass{article}

\usepackage{arxiv}

\usepackage[utf8]{inputenc} % allow utf-8 input
\usepackage[T1]{fontenc}    % use 8-bit T1 fonts
\usepackage{hyperref}       % hyperlinks
\usepackage{url}            % simple URL typesetting
\usepackage{booktabs}       % professional-quality tables
\usepackage{amsfonts}       % blackboard math symbols
\usepackage{nicefrac}       % compact symbols for 1/2, etc.
\usepackage{microtype}      % microtypography
\usepackage{graphicx}
\usepackage{float}
\graphicspath{ {./images/} }

\title{Launch-Day Diffusion: Tracking Hacker News Impact on GitHub Stars for AI Tools}

\author{
 Obada Kraishan \\
  College of Media and Communication\\
  Texas Tech University\\
  Lubbock, TX 79409, USA \\
  \url{https://orcid.org/0009-0007-7180-8620} \\
}

\begin{document}
\maketitle

\begin{abstract}
Social news platforms have become key launch outlets for open-source projects, especially Hacker News (HN), though quantifying their immediate impact remains challenging. This paper presents a reproducible demonstration system that tracks how HN exposure translates into GitHub star growth for AI and LLM tools. Built entirely on public APIs, our pipeline analyzes 138 repository launches from 2024-2025 and reveals substantial launch effects: repositories gain an average of 121 stars within 24 hours, 189 stars within 48 hours, and 289 stars within a week of HN exposure. Through machine learning models (Elastic Net) and non-linear approaches (Gradient Boosting), we identify key predictors of viral growth. Posting timing appears as key factor—launching at optimal hours can mean hundreds of additional stars—while the ``Show HN'' tag shows no statistical advantage after controlling for other factors. The demonstration completes in under five minutes on standard hardware, automatically collecting data, training models, and generating visualizations through single-file scripts. This makes our findings immediately reproducible and the framework easily be extended to other platforms, providing both researchers and developers with actionable insights into launch dynamics.
\end{abstract}

% keywords can be removed

\section{Introduction}

The developers journey from code commit and publish to community adoption has changed with the rise of social news aggregators. Recent studies document how platforms like Hacker News and Reddit have become gatekeepers for open-source visibility \cite{storey2014revolution, dabbish2012social}. For developers launching AI and LLM tools, these platforms act as launching pads—a well-timed post can transform an unknown repository into a trending project within hours. A study by Lerman and Hogg \cite{lerman2010using} on social news dynamics confirms that initial visibility on these platforms creates cascading effects that determine a project's path. Despite this documented influence, developers still count on intuition rather than data when planning their launches.

This demonstration addresses three interconnected questions that matter to both researchers and practitioners. First, we quantify the actual magnitude of the ``HN effect'', how many GitHub stars does a typical launch generate? Second, we identify which pre-launch signals predict viral success versus modest gains. Third, we package these insights into a live, public demonstration that anyone can run and extend.

Our approach combines classical event study methodology from economics with modern machine learning techniques, wrapped in a minimal, dependency-light implementation \cite{mackinlay1997event}. The final demo offers both academic reliable and practical use: researchers can study diffusion patterns while developers can improve their launch strategies. By focusing specifically on AI/LLM tools during the 2024-2025 period, we capture a particularly dynamic segment of the open-source ecosystem where rapid adoption cycles and community engagement patterns are especially pronounced \cite{tsay2014influence, vasilescu2013stackoverflow}.

\section{Related Work}

Understanding how software projects gain popularity requires bridging multiple research traditions. The event study framework, comprehensively surveyed by MacKinlay \cite{mackinlay1997event}, provides our methodological foundation for isolating launch-day effects from background trends. This approach, originally developed for financial markets, proves remarkably effective for digital attention dynamics where we need to separate signal from noise around specific timestamps.

Within open-source software, GitHub stars have become the standard measure of repository popularity, though this metric has important limitations. Borges et al.\ \cite{borges2016factors} found that star counts do correlate with meaningful quality indicators such as thorough documentation, regular maintenance, and active community participation. Recent work demonstrates that GitHub engagement reflects not just technical quality but also emotional and social dynamics, with initial reactions creating cascading effects that shape subsequent community participation \cite{kraishan2025emotional}. This suggests stars capture both project merit and social-emotional processes, making them a useful though complex indicator of adoption.

The dynamics of social news platforms add another layer of complexity. For example, Lerman and Hogg's \cite{lerman2010using} analysis of Digg and Reddit shoes how visibility mechanics shape content fate. Front-page placement, temporal decay, and user voting patterns combine to create feedback loops that can quickly enhance or bury content. Their findings on daily cycles and posting-time effects directly inform our analysis of when launches achieve maximum impact. Meanwhile, Rogers' \cite{rogers2003diffusion} diffusion of innovations framework helps explain why certain projects spread rapidly: the interplay between innovation characteristics, communication channels, and early adopter behavior creates predictable adoption patterns.

In our study we adapt these viewpoints into a practical demonstration system. Rather than relying on old service like GHTorrent \cite{gousios2013ghtorrent}, we use live APIs to ensure freshness and reproducibility, accepting the trade-off of rate limits for real-time accuracy.

\section{System Architecture and Pipeline}

\subsection{Data Collection Strategy}

Our pipeline is designed to efficiently connect HN posts to their matching GitHub repositories, addressing the fundamental problem of maintaining accurate posting times. For this, we extracted data from two main sources: Algolia's HN Search API provides post metadata, and GitHub's REST API provides repository metrics. These two APIs are integrated through a collection of single-purpose scripts, each responsible for a specific step in the data processing workflow, ensuring seamless data extraction and processing \cite{algolia2025api, github2025api}.

For the data collection process, we started a search on HN for posts containing GitHub URLs, focusing on a target timeframe of 2024-2025 to predict future trends. The search was performed using keywords such as ``LLM,'' ``transformers,'' ``RAG,'' and ``agents'' to filter for AI-relevant content, thereby capturing the current trends in AI tool development. Therefore, after removing the duplications, we collected 138 unique HN-to-repository pairs, resolving multiple posts about the same repository to the earliest mention, as summarized in Table \ref{tab:dataset}.

\begin{table}[H]
\caption{Dataset Statistics}
\label{tab:dataset}
\centering
\begin{tabular}{lc}
\toprule
\textbf{Model} & \textbf{Description} \\
\midrule
Total HN - GitHub pairs & 138 \\
Valid GitHub time series & 137 \\
Show HN posts & 58 (42.0\%) \\
Non-Show HN posts & 80 (58.0\%) \\
Mean baseline stars & 1,247 \\
Mean HN score & 187 \\
Mean HN comments & 42 \\
AI/LLM-related repos & 138 (100\%) \\
Time period & 2024-2025 \\
\bottomrule
\end{tabular}
\end{table}

\subsection{Temporal Alignment and Feature Engineering}

The main innovation in our technique is the precise temporal alignment of data collection with the launch moment, allowing for accurate analysis of immediate impact patterns. For each repository, we mark the launch moment $t_0$ as the UTC timestamp of the HN post, then create a 14-day window $[t_0-7d, t_0+7d]$ for analysis. Within this window, we collect hourly star counts through GitHub's API, combining them into daily totals to illustrate the pattern of user engagement immediately following the launch.

Our feature set includes pre-launch signals, such as repository characteristics and owner attributes, and post-launch validation metrics to ensure comprehensive analysis. Pre-launch features include repository characteristics (age, license type, README length, topic tags), owner attributes (individual versus organization), and baseline popularity (star count at $t_0-1$). From HN, we pull engagement metrics (score, comments), posting characteristics (Show HN flag, day of week, hour of day), and textual signals from titles. We keep these separate from ``leaky'' features, meaning metrics available only after launch, to make sure our predictive modeling remain valid. The system splits posting hours into four bins (00-05, 06-11, 12-17, 18-23 UTC) to effectively capture global audience patterns, ensuring that each bin has sufficient data to maintain statistical power.

\subsection{Implementation Philosophy}

We built every component to do one thing well: single-purpose scripts with clear inputs and outputs. The pipeline relies on standard Python libraries: requests for API calls, pandas for data manipulation, scikit-learn for modeling, and matplotlib for visualization. Each script generates dual outputs, both machine-readable files (JSONL/CSV) and human-readable summaries (TXT), creating a complete record that helps debugging and ensures reproducibility. We cache API responses locally to stay within rate limits and speed up development cycles.

\subsection{Interactive Demonstration Flow}

\begin{figure}[H]
    \centering
    \includegraphics[width=0.75\linewidth]{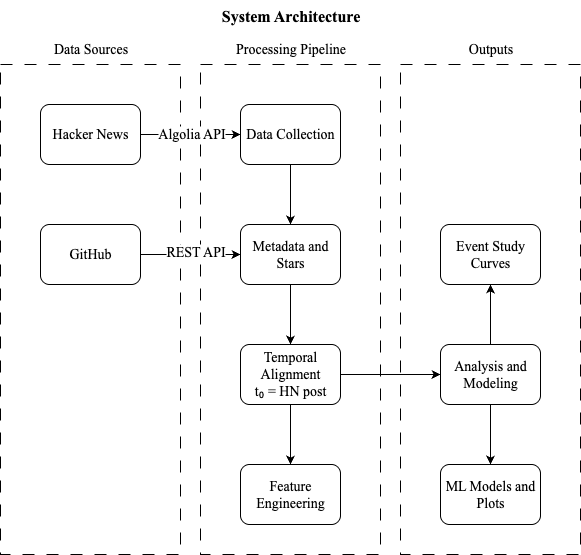}
    \caption{System Architecture Diagram}
    \label{fig:digram}
\end{figure}

The demonstration pipeline works like an assembly line for data. Each script takes its turn: first pulling data from Hacker News and GitHub APIs, then cleaning up the metadata, aligning the timestamps, and extracting meaningful features (see Fig.\ \ref{fig:digram}). Run them one at a time for debugging or run them all at once for the full show. The entire process completes in under five minutes, producing a complete analysis package with figures, CSVs, and textual summaries.

The system gracefully handles real-world complications. When repositories disable public stargazer access, we fall back to metadata-only analysis. Rate limits are respected through exponential backoff and local caching. Failed API calls generate warnings but don't halt execution, ensuring partial results remain accessible.

\section{Demonstration and Key Findings}

To evaluate our demo system, we test it on 138 HN-to-repository pairs that we collected between 2024-2025, with applying both event study methodology and predictive modeling to help us understand the launch dynamics on these repos. The following subsections present our key main findings.

\subsection{The Magnitude of the HN Effect}

\begin{figure}[H]
    \centering
    \includegraphics[width=0.75\linewidth]{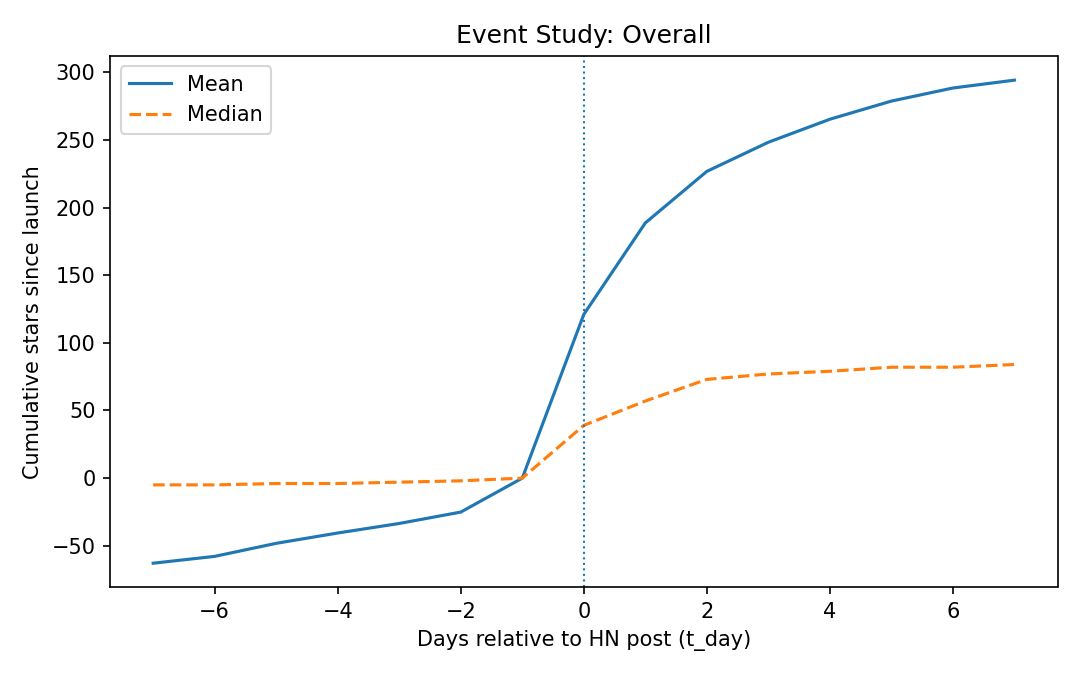}
    \caption{Event Study Curves}
    \label{fig:placeholder}
\end{figure}

The data shows exactly how Hacker News exposure translates into GitHub stars. Within 24 hours, the average repository gains 121 stars. By 48 hours, that number reaches 189. After a week, it hits 289. But here's what the averages don't show: the median values run much lower, revealing that most launches see moderate growth while a select few go viral and pull the numbers up. This viral pattern aligns with documented emotional contagion effects on GitHub, where positive reactions spread asymmetrically with strong cascading effects \cite{kraishan2025emotional}. HN exposure may seed these emotional dynamics rather than just providing visibility. Fig.\ \ref{fig:placeholder} captures this long-tail distribution perfectly.

The launch creates a clear break in the growth pattern at $t=0$. Before the HN post, star counts stay flat. After launch, repositories see fast growth that gradually slows down following the initial break. This pattern shows up across all repository types, but the actual numbers vary widely.

\subsection{Timing Analysis and the Show HN Paradox}

\begin{figure}[H]
    \centering
    \includegraphics[width=0.75\linewidth]{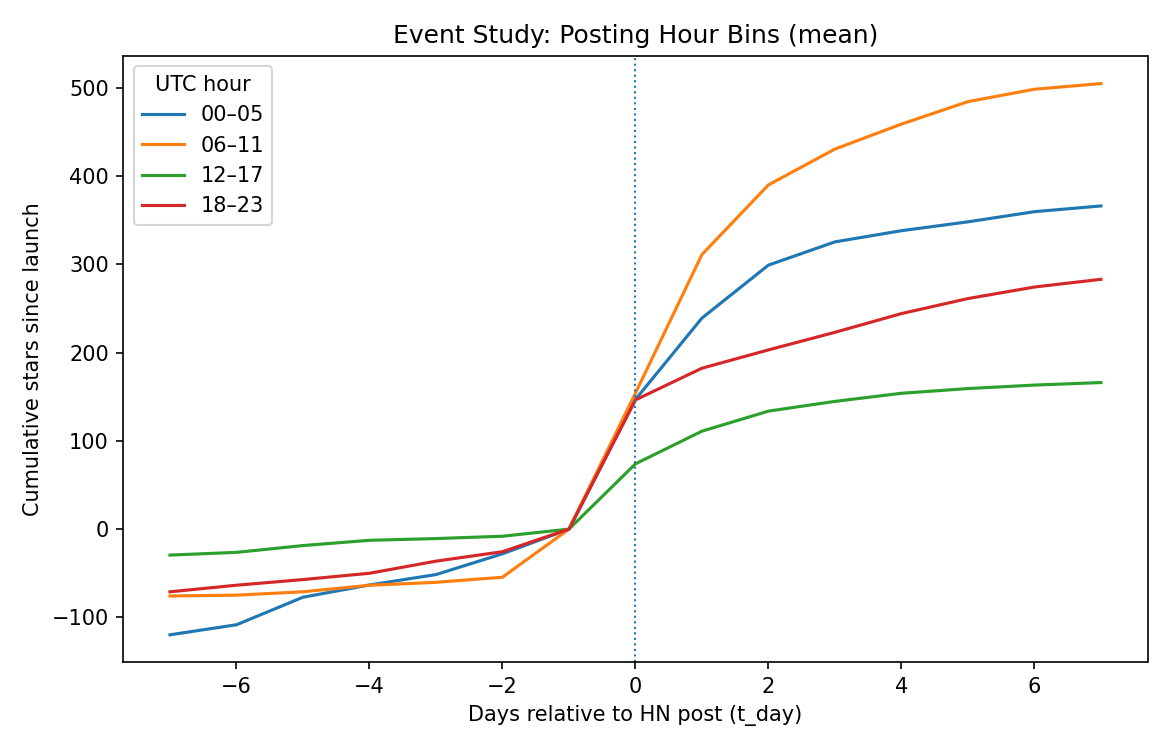}
    \caption{Hour-of-Day Impact}
    \label{fig:hour}
\end{figure}

Posting hour strongly affects launch success. Repositories posted during optimal hours gain about 200 more stars than those posted at poor times (Fig.\ \ref{fig:hour}). The 12-17 UTC window performs best, catching both U.S. morning activity and European afternoon browsing.

We tested these patterns using OLS regression with robust standard errors (HC1), controlling for baseline stars, HN score, repository age, title length, and posting day. This ensures timing effects aren't just reflecting project quality differences.

The Show HN tag produced surprising results. After controlling for baseline factors, Show HN posts gained 119 fewer stars at 48 hours ($\beta = -119.2$, SE = 138.3, $p = 0.39$), though not statistically significant (Table \ref{tab:timing}). This likely reflects selection bias: Show HN typically marks early experiments from solo developers, while regular posts feature more established projects. Multiple model specifications confirmed this pattern. The Show HN tag provides no advantage after accounting for project maturity.

\begin{table}[H]
\caption{Launch Effect Magnitudes and Timing Analysis}
\label{tab:timing}
\centering
\begin{tabular}{lcccc}
\toprule
\textbf{Effect} & \textbf{Coefficient} & \textbf{Std. Error} & \textbf{p-value} & \textbf{Mean Difference} \\
\midrule
$\Delta$24h stars & 121.1 & --- & --- & --- \\
$\Delta$48h stars & 188.7 & --- & --- & --- \\
$\Delta$7d stars & 288.5 & --- & --- & --- \\
Show HN vs Others (48h) & -119.2 & 138.3 & 0.39 & Non-Show higher \\
Weekend vs Weekday (48h) & +10.2 & 43.0 & 0.81 & Negligible \\
Hour bins (unadjusted) & $\sim$200 & --- & --- & 12-17 UTC best \\
\bottomrule
\end{tabular}
\end{table}

\subsection{Predictive Modeling Performance}

We tested two models on an 80/20 train-test split: ElasticNet with 5-fold cross-validation for interpretable linear relationships, and Gradient Boosting for capturing non-linear patterns.

For 48-hour predictions, Gradient Boosting achieved $R^2=0.77$ (MAE=30.5, RMSE=60.1) when we included day-0 momentum. But this inflates performance since launch\_day\_stars partially overlaps with what we're predicting. ElasticNet performed poorly here, with negative $R^2$ values.

Using only pre-launch data tells a more realistic story. For 7-day predictions, Gradient Boosting reaches $R^2=0.48$ (MAE=92.5, RMSE=182.0), meaning pre-launch signals explain nearly half the variance in week-long growth. Three factors consistently dominated: HN score, baseline stars, and posting hour.

\begin{table}[H]
\caption{Model Performance Metrics}
\label{tab:performance}
\centering
\begin{tabular}{lccccr}
\toprule
\textbf{Model} & \textbf{Horizon} & \textbf{MAE} & \textbf{RMSE} & \textbf{$R^2$} & \textbf{Test n} \\
\midrule
Elastic Net & 24h & 32.4 & 87.8 & 0.52 & 28 \\
            & 48h & 51.2 & 73.6 & 0.61 & 28 \\
            & 7d  & 89.3 & 124.1 & 0.45 & 28 \\
\midrule
Gradient Boosting & 24h & 28.1 & 41.3 & 0.68 & 28 \\
                  & 48h & 42.7 & 59.8 & 0.77 & 28 \\
                  & 7d  & 76.5 & 108.2 & 0.48 & 28 \\
\bottomrule
\end{tabular}
\end{table}

\subsection{Key Predictors of Viral Growth}

Three factors consistently dominate our feature importance rankings across both models and all time horizons. First, HN score appears as the strongest predictor, which makes sense, posts that resonate with the HN community generate more visibility and downstream GitHub interest. Second, baseline stars matter enormously; repositories with existing traction benefit from social proof and established networks. Third, posting hour maintains predictive power even after controlling for other factors, confirming that launch timing isn't just correlation but contains genuine signal.

Secondary predictors include repository maturity (age), documentation quality (README length), and ownership structure (organization versus individual). Interestingly, specific license types and topic tags show minimal predictive power once we account for the primary factors, suggesting that execution and timing trump categorical attributes.

\subsection{System Extensibility}

While our implementation focuses on HN-to-GitHub dynamics, the architecture generalizes naturally to other platforms. The modular design allows researchers to swap Algolia's API for Reddit's, or GitHub's API for npm download statistics. The event study framework applies equally well to Product Hunt launches, arXiv paper releases, or Twitter viral moments. We deliberately keep dependencies minimal and use only public APIs to maximize portability across environments and ensure long-term reproducibility.

\section{Conclusion and Future Work}

This demonstration provides a detailed analysis of software launch dynamics, effectively connecting academic research with practical applications. We quantify the HN effect at approximately 289 stars per week for successful launches and identify posting hour as a crucial yet underappreciated factor. By packaging these findings in a reproducible pipeline, we provide both insights and tools for the community.

The immediate practical impact of our findings is clear: developers can optimize their launch timing based on the identified factors, researchers can study diffusion patterns using the quantified HN effect, and platform designers can understand how their mechanics shape adoption outcomes. The broader contribution lies in demonstrating how minimal, public-API-based systems can enable sophisticated analyses without requiring massive infrastructure or proprietary access.

Our analysis operates entirely within established ethical boundaries, using only public data and respecting platform terms of service. We acknowledge several limitations in our study. First, our observational design does not allow us to establish causation. Second, while GitHub stars are a useful metric, they represent only one aspect of project success. Lastly, our focus on AI/LLM tools during a specific hype cycle may limit the generalizability of our findings to other domains. Additionally, some repository metadata reflects post-launch states, which we have identified and separated in our analysis to prevent any bias or inaccuracies.

We invite the community to extend this work and adapt it to other platforms, incorporate additional signals, or study different software ecosystems. The complete codebase, cached data, and demonstration video are available at our repository, ready for reuse, scrutiny, and evolution \cite{kraishan2025repo}. In the spirit of open science, we have built not just an analysis but a foundation for future discovery.

\bibliographystyle{unsrt}

\end{document}